# Internal friction and Jahn-Teller effect in the charge-ordered $La_{1-x}Ca_xMnO_3$ ($0.5 \le x \le 0.87$)

R. K. Zheng, R. X. Huang, A. N. Tang, G. Li, X. G. Li [a)]

Structure Research Laboratory, Department of Materials Science and Engineering, University of Science and Technology of China, Anhui, Hefei, 230026, P. R. China

J. N. Wei, J. P. Shui

Laboratory of Internal Friction and Defects in Solids, Institute of Solid State Physics, Chinese Academy of Science, Anhui, Hefei, 230031, P. R. China

Z. Yao

Department of Physics, University of Texas at Austin, Austin, Texas, 78712

The Jahn-Teller effect in the charge-ordered (CO) state for $La_{1-x}Ca_xMnO_3$ ($0.5 \le x \le 0.87$) was studied by measuring the low-temperature powder x-ray diffraction, internal friction, and shear modulus. We find that the electron-lattice interaction with the static Jahn-Teller distortion is the strongest near $x \approx 0.75$ in the CO state. It was particularly observed that a crossover of the Jahn-Teller vibration mode from $Q_2$ to $Q_3$ near $x$=0.75 induces crossovers of the crystal structure from tetragonally compressed to tetragonally elongated orthorhombic, and of the magnetic structure from CE-type to C-type near $x$=0.75. The experimental results give strong evidence that the Jahn-Teller effect not only plays a key role in stabilizing the CO state, but also determines the magnetic and crystal structures in the CO state for $La_{1-x}Ca_xMnO_3$.

[a)] Corresponding author, Electronic mail: lixg@ustc.edu.cn

It is well known that the $La_{1-x}Ca_xMnO_3$ ($0.5 \le x \le 0.87$) manganites show charge, spin, and/or orbital orderings below the charge ordering transition temperature $T_{CO}$ [1], and much efforts have been devoted to this system to disclose the microscopic origin of the charge-ordered (CO) state [2,3]. It has been suggested that when the long-range Coulomb interaction and/or a strong electron-lattice interaction with the Jahn-Teller (JT) distortion overcomes the kinetic energy of $e_g$ electrons, real-space charge and concomitant spin and/or orbital orderings occur throughout the crystal structure. Despite many investigations have been done on this aspect, the main driving force of the CO state being the long-range Coulomb interaction or the electron-lattice interaction with the JT effect or both is still a subject of discussion [3,4]. Recent experimental observation of "wigner-crystal" CO state from transmission electron microscopy, synchrotron x-ray and neutron diffractions on $La_{0.33}Ca_{0.67}MnO_3$ [2,5,6] demonstrates that the long-range Coulomb interaction might be the main driving force of the CO state, and indeed, some theoretical calculations [7,8] support the Coulombic model. However, if CO state was mainly due to long-range Coulomb interaction, the $z$-axis stacking of charges [9] and the observed "bi-stripe" CO state [10] which are both energetically penalized by the nearest-neighbor Coulomb repulsion $V_{NN}$, therefore, can not be fully understood based on the Coulombic model. This shows that other ingredients, especially the electron-lattice interaction with the JT effect, are needed to understand the formation of CO state. The theoretical calculations in Refs. [3,4,11-16] have shown that the JT effect not only stabilizes the CO state, but also strongly affects the magnetic structures. The relative importance of the long-range Coulomb interaction or the electron-lattice interaction with the JT effect in the CO state is yet to be experimentally resolved. In this paper, we experimentally *highlight* the prominent role of the JT effect in the CO state. The relations among the JT vibration mode, the crystal and magnetic structures were also studied.

The polycrystalline $La_{1-x}Ca_xMnO_3$ ($x$=0.5, 0.55, 0.6, 0.65, 0.7, 0.75, 0.8, 0.83, 0.85, 0.87) samples in the form of 60mm×5mm×2mm rectangular bars were synthesized via a coprecipitation method. The orthorhombic lattice parameters of $La_{1-x}Ca_xMnO_3$ were determined by a Japan MXP18AHF powder x-ray diffractometer using Cu $K\alpha$ radiation ($\lambda$=1.54056 Å) at various temperatures. Low-frequency internal friction ($Q^{-1}$) and shear modulus ($G$) were measured on a multi-functional internal friction apparatus using the forced-vibration method with five different vibration frequencies of 0.1, 0.72, 1.00, 1.88 and 5.11 Hz during the warm-up from 120 K to 475 K at a heating rate of 2.0 ºC/min.

Fig. 1 shows the temperature dependence of the shear modulus ($G$) at five different frequencies for $La_{1-x}Ca_xMnO_3$ ($0.5 \le x \le 0.87$). For all the samples, upon cooling from a high temperature the $G$ softens conspicuously above $T_{CO}$ and stiffens dramatically below $T_{CO}$, and an internal friction peak also appears near $T_{CO}$. Only typical internal friction peaks ($Q^{-1}$) for $x$=0.55, 0.75, and 0.8 are shown in the insets of Fig. 1. We note that the softening and stiffening of the shear moduli and the internal friction peak positions are vibration frequency independent. Nevertheless, the height of the internal friction peak decreases with increasing measuring frequencies. These characteristics of $G$ and $Q^{-1}$ suggest that the shear moduli and internal friction are closely related to the structural changes upon the CO transition, and no thermally activated relaxation process is involved in the CO transition. Early synchrotron x-ray and neutron diffractions have confirmed that the structural (*i.e.* lattice parameters) changes near $T_{CO}$ are microscopically due to the development of the static JT type lattice distortion upon the formation of CO state [2,6]. The stiffening of the shear modulus below $T_{CO}$, therefore, corresponds to the development of the static cooperative JT type lattice distortion in the CO state. Hence, it can be regarded that the magnitude of the stiffening of the shear modulus ($\Delta G/G$) directly reflects the magnitude of the static JT distortion in the CO state.

Fig. 2 (a) shows the relative changes of the shear modulus $\Delta G/G$ in the CO state as a function of doping level $x$. The $\Delta G/G$ increases almost smoothly with increasing $x$ from 0.5 to 0.75, then decreases for $x$>0.75. As already mentioned above the $\Delta G/G$ can be a measure of the magnitude of the static JT distortion in the CO state. The variation of $\Delta G/G$ with $x$, thus, implies that the magnitude of the static JT distortion and its associated effective strength of electron-lattice interaction strongly depend on the doping level $x$, and are the strongest at $x$=0.75. As is well known, by distorting the $MnO_6$ octahedra, the degeneracy of the doublet $e_g$ levels is lifted, and the JT energy gain ($E_{JT}$) which is linear with the magnitude of the distortion is of 1/4 of the splitting of the $e_g$ levels [17]. Therefore, in the JT effect viewpoint, the $E_{JT}$ is the largest for $x$=0.75, as determined from the $\Delta G/G$~$x$ relationship shown in Fig, 2(a). The angle-resolved photoemission experiments on parent compound $LaMnO_3$ indicate that the JT energy gain due to the formation of cooperative static JT distortion with coherent orbital ordering is of 0.25eV [17]. By comparing the $\Delta G/G$ for $0.5 \leq x \leq 0.87$ with that of $LaMnO_3$ ($\Delta G/G$=9%) (see Fig. 3), it is qualitatively estimated that the $E_{JT}$ for $x \approx 0.75$ is much larger than 0.25eV. Although our experimental results can not determine the effect of the long-range

Coulomb interaction in the CO state, the large energy gain due to the strong static JT distortion contributes greatly to stabilizing the CO state, especially for $x \approx 0.75$. This scenario is consistent with the theoretical calculations in Refs [3,4,11-16] that a relatively strong electron-lattice interaction with the JT distortion can stabilize the spin, charge, and orbital orderings. On the other hand, one may expect that the strength of the long-range Coulomb interaction between the $e_g$ electrons with different orbitals for $La_{1-x}Ca_xMnO_3$ ($0.5 \leq x \leq 0.87$) becomes increasingly smaller with increasing $x$ (especially for $x > 0.75$) because of the decrease of the $e_g$ electronic density with increasing $x$. These scenarios and our recent experimental observation [18] that the CO state is the most robust near $x \approx 0.75$, qualitatively suggest that the electron-lattice interaction with the JT distortion is very likely the main driving force of the CO state.

To further characterize the JT effect in the CO state, we show, in Fig. 2(b), the ratio of $(b/\sqrt{2})/a$ (or almost equivalently the ratio of the apical to equatorial Mn-O bond length [19]) at 30 K as a function of doping level $x$ for $La_{1-x}Ca_xMnO_3$. With increasing $x$ from 0.5 to 0.75, the $(b/\sqrt{2})/a$ ratios increase smoothly, nevertheless, they are smaller than 1, indicating that the $b$-axis of the orthorhombic lattice is compressed. With further increase of $x$ the $(b/\sqrt{2})/a$ ratios increase abruptly around $x=0.75$, and are larger than 1 for $x > 0.75$, suggesting an elongation of the $b$-axis with respect to that of ideal cubic lattice. As is known, the degeneracy of the doublet $e_g$ levels is lifted by tetragonal distortions of the $MnO_6$ octahedra with either $Q_2$ or $Q_3$ mode JT distortion. This $b$-axis compressed and elongated lattices are obviously a manifestation of different type JT vibration modes that are $Q_2$ for $0.5 \leq x \leq 0.75$ and $Q_3$ for $x > 0.75$, and moreover, the abrupt increase of $(b/\sqrt{2})/a$ suggests that there is a crossover of the JT vibration mode from $Q_2$ to $Q_3$ around $x$=0.75 in the CO state.

The interplay between JT distortion and magnetic structures has been theoretically studied by many groups, and all these calculations show that the JT distortion with the orbital ordering strongly affect the magnetic structures [11-16]. We note that the pioneering neutron diffraction measurements on $La_{1-x}Ca_xMnO_3$ ($0.5 \le x \le 1.0$) by Wollan and Koehler [9] have shown that there is a crossover of magnetic structure from CE-type to C-type near $x$=0.75 in the CO state. This crossover of magnetic structure seemed to be coupled with the crossover of the JT vibration mode from $Q_2$ to $Q_3$ around $x$=0.75. It's worth mentioning here, based on Goodenough's theory of semicovalent exchange, the microscopic origin of the CE-type magnetic structure in half-doped manganites (*i.e.* $x$=0.5) is soundly illustrated in terms of a shear mode lattice deformation driven by $Q_2$ mode JT distortion [13]. On the other hand, Konishi *et al.*'s [20] recent study on single crystals of $Nd_{1-x}Sr_xMnO_3$ ($x>0.5$) and epitaxial thin films of $La_{1-x}Sr_xMnO_3$ grown on $LaAlO_3$ and $SrTiO_3$ with the ratio of lattice parameters $c/a>1$ and $c/a<1$ show that the magnetic structures for those with $c/a>1$ is ubiquitously C-type, and it is increasingly stabilized with increasing $c/a$ ratio (*i.e.* with increasing apically elongated JT distortion) [19,20]. Keeping these features in mind, one may conclude that the C-type magnetic structure for $La_{1-x}Ca_xMnO_3$ with $x>0.75$ [9] is directly driven by a apically elongated JT distortion (*i.e.* $Q_3$ mode) accompanied with the ($3z^2$-$r^2$) orbital ordering, and either the CE-type or C-type magnetic structure can be increasingly stabilized with increasing $Q_2$ or $Q_3$ mode JT distortion. Thus, we can safely conclude that the crossover of magnetic structure from CE-type to C-type near $x$=0.75 microscopically originates from the crossover of the JT vibration mode from $Q_2$ to $Q_3$. As mentioned above the strong JT distortion with the orbital ordering significantly stabilize the CE- or C-type magnetic structure, the CE- or C-type magnetic ordering, especially for 0.65≤$x$≤0.83 having large JT distortion, is expected to be very stable, and thus to be unsusceptible to an external magnetic field. Based on this picture, it is expected that the CO state and the charge transport properties for 0.65≤$x$≤0.83 will not be easily altered under an external magnetic field, which agrees with the very weak magnetic field dependence of the resistivity up to 14 Telsa for $La_{1-x}Ca_xMnO_3$ (0.65≤$x$≤0.83) [18].

In summary, the experimental results demonstrate that the anomalies in the charge, spin, lattice, and orbital degree of freedoms at (or near) $x$=6/8 is one of the most important anomalies besides those observed at $x$=N/8 (N=1, 3, 4, 5, 7) for $La_{1-x}Ca_xMnO_3$. The dramatic stiffening of the shear modulus implies there presents a strong JT type lattice distortion in the CO state, especially for $x \approx 0.75$, which greatly stabilizes the CO state. The magnetic and crystal structures in the CO state are dependent on different types of JT vibration modes which changes from $Q_2$ mode to $Q_3$ mode near $x$=0.75 in the CO state, which strongly implies that the CO state has its origin of the cooperative nature of the JT type lattice distortion.

This work was supported by the Chinese National Nature Science Fund, the Ministry of Science and Technology of China.

## Figure and Table captions

Fig. 1. Temperature dependence of the shear modulus and internal friction measured with five different vibration frequencies for $La_{1-x}Ca_xMnO_3$ ($0.5 \le x \le 0.87$). The shear modulus minimum corresponds to $T_{CO}$.

Fig. 2. The relative change of the shear modulus *ΔG/G vs.* doping level *x* (a), and the ratio of lattice parameters $(b/\sqrt{2})/a$ *vs.* doping level *x* at 30 K (b). Insets in (a) show the $Q_2$ and $Q_3$ mode JT distortion for $x \le 0.75$ and $x > 0.75$, respectively. Insets in (b) show the tetragonally compressed and elongated primitive perovskite orthorhombic lattice for $x \le 0.75$ and $x > 0.75$, respectively.

Fig. 3. Temperature dependence of the shear modulus measured with four different vibration frequencies for $LaMnO_3$.

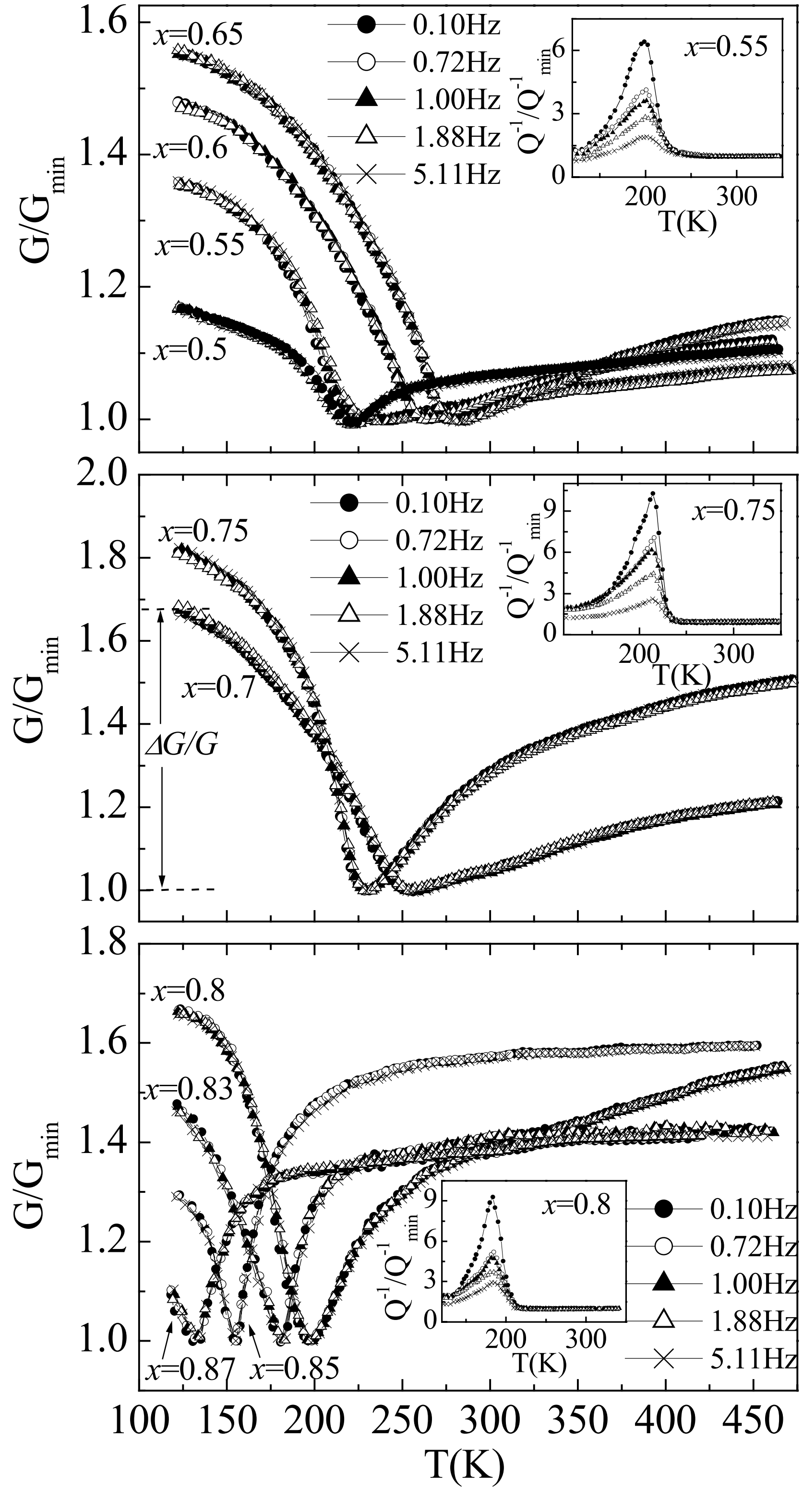

Fig. 1 By R. K. Zheng *et al.*

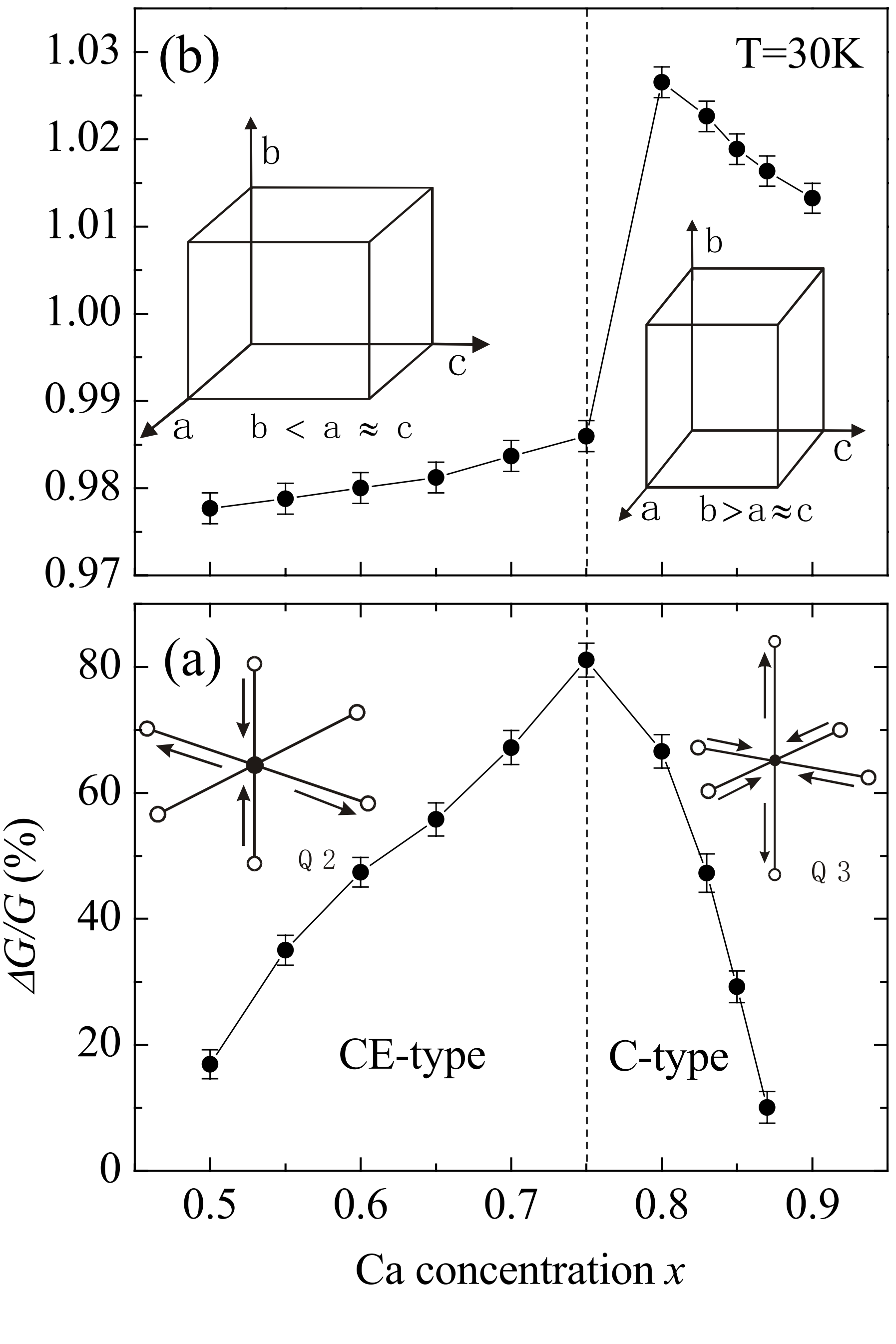

Fig. 2 By R. K. Zheng *et al*.

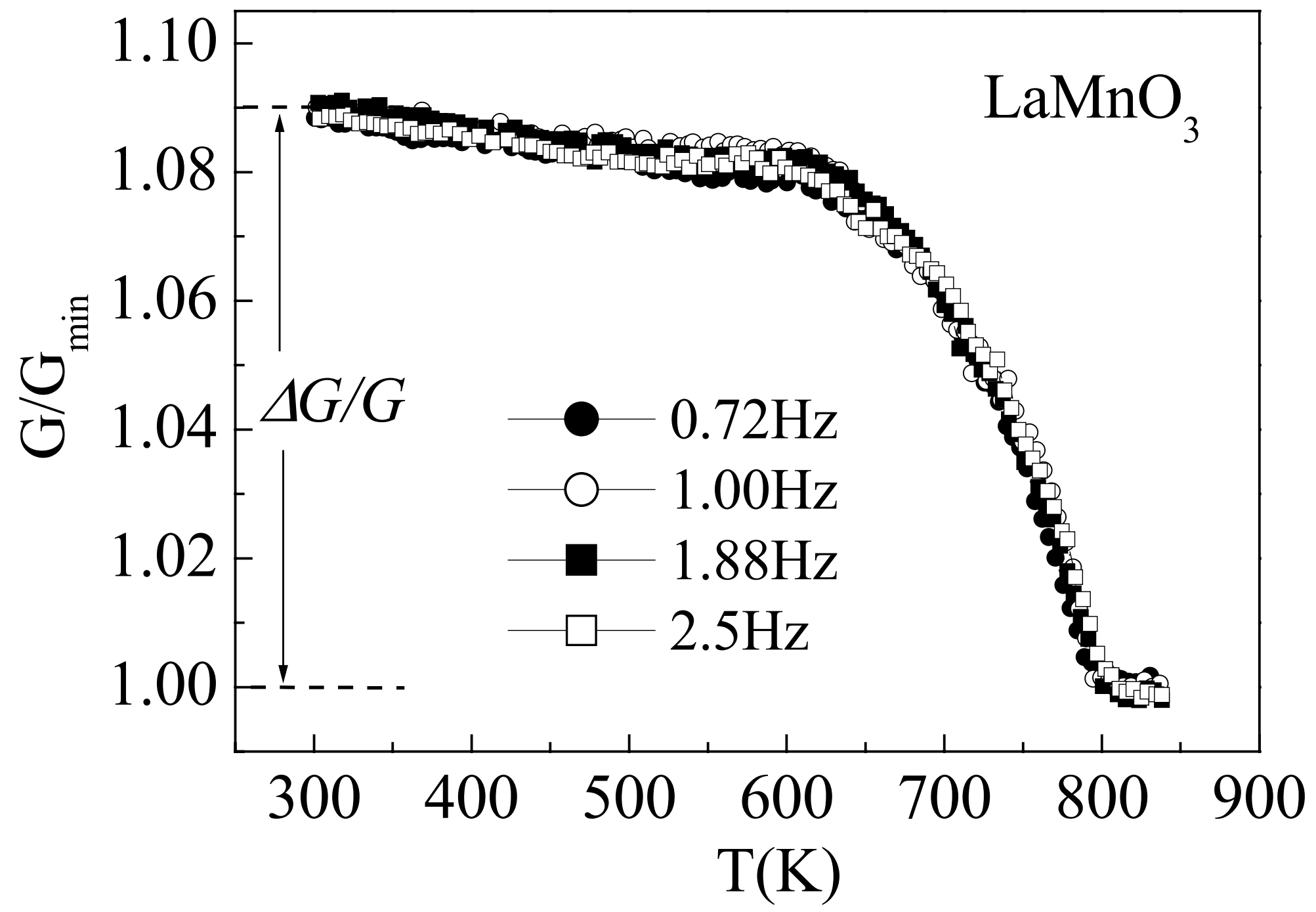

Fig. 3 By R. K. Zheng *et al*.